\documentclass[twocolumn,pre,superscriptaddress,showpacs]{revtex4}

\usepackage[german,english]{babel}
\usepackage{amssymb}
\usepackage{amsfonts}

\begin{document}

\title{L{\'e}vy flights in a steep potential well}
\date{\today}

\author{Aleksei V. Chechkin}
\affiliation{Institute for Theoretical Physics NSC KIPT,
Akademicheskaya st.1, 61108 Kharkov, Ukraine}
\author{Vsevolod Yu. Gonchar}
\affiliation{Institute for Theoretical Physics NSC KIPT,
Akademicheskaya st.1, 61108 Kharkov, Ukraine}
\author{Joseph Klafter}
\affiliation{School of Chemistry, Tel Aviv University, 69978 Tel Aviv Israel}
\author{Ralf Metzler}
\affiliation{NORDITA, Blegdamsvej 17, DK-2100 Copenhagen \O, Denmark}
\thanks{Corresponding author. Tel: +45 353 25507. Fax: +45 353 89 157.
E-mail: metz@nordita.dk}
\author{Leonid V. Tanatarov}
\affiliation{Institute for Theoretical Physics NSC KIPT,
Akademicheskaya st.1, 61108 Kharkov, Ukraine}

\begin{abstract}
L{\'e}vy flights in steeper than harmonic potentials have been shown to
exhibit finite variance and a critical time at which a bifurcation from
an initial mono-modal to a terminal bimodal distribution occurs (Chechkin
et al., Phys. Rev. E {\bf 67}, 010102(R) (2003)). In this paper, we
present a detailed study of L{\'e}vy flights in potentials of the type
$U(x)\propto|x|^c$ with $c>2$. Apart from the bifurcation into bimodality,
we find the interesting result that for $c>4$ a trimodal transient exists
due to the temporal overlap between the decay of the central peak around
the initial $\delta$-condition and the  building up of the two emerging
side-peaks, which are characteristic
for the stationary state. Thus, for certain system parameters there exists
a transient tri-modal distribution of the L{\'e}vy flight. These properties
of LFs in external potentials of the power-law type can be represented by
certain phase diagrams. We also present details about the
proof of multi-modality and the numerical procedures to establish the
probability distribution of the process.\\[0.4cm]
Running head: L{\'e}vy flights in a steep potential well
\end{abstract}


\maketitle

\section{Introduction}

L{\'e}vy flights (LFs) are stochastic, Markov processes, which differ
from regular Brownian motion by the occurrence of extremely long jumps,
whose length is distributed according to a L{\'e}vy stable law with the
long tail $\sim|x|^{-1-\alpha}$, such that its second moment diverges
\cite{hughes,bouchaud,levy,gnedenko,pt}. This property strongly contrasts
the classical
Gaussian description of diffusion processes which possess finite moments
of any given order \cite{chandra,vankampen}. Given their Markov nature,
the divergence of the variance of an LF might disqualify them as physically
meaningful model processes for diffusing particles with a finite mass.
Yet, LFs have important applications to processes, in which no finite
velocity is required, such as in the energy diffusion in single molecule
spectroscopy \cite{zukla}. An impressive experimental evidence of L{\'e}vy
processes was reported by the group of Walther in the study of the position
of a single ion in a one-dimensional optical lattice, in which diverging
fluctuations could be observed in the kinetic energy \cite{walther}. From
a phenomenological point of view, LFs have been used to describe the
dynamics observed in plasmas \cite{chechkin1} or in molecular collisions
\cite{galgani}. They have also been successfully applied to describe the
statistics encountered in the spatial gazing patterns of bacteria
\cite{levandowski}, albatross birds \cite{stanley}, or even spidermonkeys
\cite{spider}. Probably the earliest application of LFs, however, may be
in the modelling of financial markets \cite{mandelbrot}. LFs were shown
to give rise to surprisingly rich band structures in periodic potentials
\cite{geisel}. Reverse engineering methods have been developed to
construct a Langevin system with L{\'e}vy noise to produce a pre-defined
steady state \cite{ido}. We also note that the stationary state of LFs in
a confining external potential is expected to approach to the one of
L{\'e}vy walks for long times \cite{klablushle,some} (the
spatiotemporally coupled version of LFs which have converging moments
of any order), and therefore the more straightforward determination of
the solution for LFs can be used to gain insight into L{\'e}vy walks.

The theory of homogeneous LFs is pretty well understood. Thus, LFs in the
continuum limit can, inter alia, be described by continuous time random
walks with long-tailed, asymptotic power-law jump length distributions
\cite{klablushle}, or alternatively by a Langevin equation with
$\delta$-correlated, L{\'e}vy noise \cite{seshadri,fogedby,honkonen}. Both
descriptions can be mapped onto the (space-) fractional diffusion equation
\cite{fogedby,chechkin0,peseckis,report,pt1}
\begin{equation}
\label{fde}
\frac{\partial f}{\partial t}=D\frac{\partial^{\alpha}}{\partial 
|x|^{\alpha}}f(x,t),
\end{equation}
where the fractional Riesz operator $\partial^{\alpha}/\partial |x|^{\alpha}$
is most easily defined in terms of its Fourier transform \cite{samko}
\begin{equation}
\label{riesz}
\int_{-\infty}^{\infty}e^{ikx}\frac{\partial^{\alpha}}{\partial
|x|^{\alpha}}f(x,t)dx\equiv -|k|^{\alpha}f(k,t);
\end{equation}
here, $0<\alpha\le 2$. In the following we restrict our analysis
to the range $1\le\alpha\le 2$. In the limit $\alpha=2$, naturally,
equation (\ref{fde}) reduces to the classical diffusion equation, which 
described Gaussian transport. From the Fourier transform of equation
(\ref{fde}), $\partial f/\partial t=-|k|^{\alpha}f(k,t)$, we conclude that
the characteristic function is
\begin{equation}
\label{levychar}
f(k,t)=\exp\left(-Dt|k|^{\alpha}\right),
\end{equation}
i.e., exactly the characteristic function of a symmetric L{\'e}vy stable
law of index $\alpha$ \cite{levy,gnedenko}. In position space, expression
(\ref{levychar}) can be represented exactly in terms of Fox $H$-functions
\cite{schneider,report}, but for our purposes in what follows, it is
enough to remember that for large $|x|$, we find $f(x,t)\sim Dt/
|x|^{1+\alpha}$, such that the variance diverges: $\langle x^2(t)\rangle=
\infty$. Also the general formulation of LFs in external potentials and
in phase space seems well founded, in terms of fractional Klein-Kramers
equations; see, e.g., references \cite{peseckis,kusnezov,lutz,gcke}.

However, much less is known about the actual behaviour of LFs in external
potentials $U(x)$, i.e., about the properties of the probability density
function (PDF) $f(x,t)$ in the
presence of such a $U(x)$. One of the few solved examples is the one of
LFs in a harmonic potential \cite{sune}, whose solution always follows
the same stable law of index $\alpha$, like in the regular Ornstein-Uhlenbeck
process the PDF always stays Gaussian. In general, LFs in the presence of an
external potential field in the overdamped case are governed by the (space-)
fractional Fokker-Planck (or Einstein-Smoluchowski) equation
\cite{peseckis,fogedby,chechkin0,report}
\begin{equation}
\label{ffpe}
\frac{\partial f}{\partial t}=\left(\frac{\partial}{\partial x}\frac{
U'(x)}{m\eta}+D\frac{\partial^{\alpha}}{\partial |x|^{\alpha}}\right)f(x,t),
\end{equation}
which has the characteristic property that the drift term enters with the
usual first order derivative and thus preserves its additive quality such
that for a constant force $F(x)=Vm\eta$ the solution is given by the
drift-free PDF taken at the similarity variable $x-Vt$, and for a general
external potential the stationary solution differs from the Boltzmann
distribution. This latter property is in
contrast to an alternative LF model suggested in reference \cite{sokolov}.

An a priori unexpected behaviour of LFs was obtained recently as solution
of the fractional Fokker-Planck equation (\ref{ffpe}), namely the occurrence
of bimodal solutions (PDFs with two maxima) and the finiteness
of the second moment in the presence of superharmonic potentials of the form
\cite{chechkin,chechkin2}
\begin{equation}
\label{superharm}
U(x)=\frac{ax^{2m+2}}{2m+2}, \quad m=1,2,.\ldots
\end{equation}
Note that the amplitude $a>0$ has dimension $[a]={\rm g}\cdot{\rm cm}^{
-2m}/{\rm sec}^{2}$. Thus, for a potential of the form $U(x)=ax^2+bx^4$
with $a,b>0$, a turnover can be tuned from the properties of the
solution of the harmonic problem (mono-modal, diverging variance) to
a finite variance and bimodal solution by varying the ratio $b:a$
\cite{chechkin2}. If such bimodality occurs, it results from a
bifurcation at a critical time $t_c$ \cite{chechkin2}.
A typical result is shown in figure \ref{figInt1}, for the quartic case
$m=1$ and L{\'e}vy index $\alpha=1.2$: from an initial $\delta$-peak,
eventually a bimodal distribution emerges. Note that we use dimensionless
quantities in the plots, as introduced below. The location of the global
maximum/maxima is displayed in figure \ref{figInt2}, where the bifurcation
is a distinct mark. In the same figure, the value of the PDF $f$ at the 
newly emerging humps is compared to the value at the origin.

In what follows, we briefly review the description of LFs in external fields
in terms of the Langevin equation with white L{\'e}vy noise and the
space-fractional Fokker-Planck equation, as well as the properties of LFs in
superharmonic potentials; and we present two proofs, for both the finite
variance in these
cases and the existence of multi-modality. We then proceed to show that for
potentials steeper than (\ref{superharm}) with $m=1$, even a tri-modal PDF
can be obtained. In two phase diagrams, we
can classify the existence of the different $n$-modal states in dependence
of the potential exponent $c$ and the L{\'e}vy index $\alpha$; and the
critical bifurcations between different $n$-modal domains as a function
of time $t$ and potential exponent $c$. In the appendix, we discuss
the numerical methods from which the PDFs are obtained.

\section{Starting equations}

In this section, we formulate the dynamical description of LFs on the
stochastic differential (Langevin equation) and the deterministic
(fractional Fokker-Planck equation) levels. For the latter, we also
discuss the corresponding form in Fourier space.

\subsection{Starting equations in real space}

\subsubsection{The Langevin equation with L{\'e}vy noise}

On the level of the stochastic description, our starting point is the
overdamped Langevin equation \cite{seshadri,fogedby,rem}
\begin{equation}
\label{langevin}
\frac{dx}{dt}=\frac{F(x)}{M\gamma}+Y_{\alpha}(t),
\end{equation}
where $F=-dU/dx$ is an external force with potential
\begin{equation}
\label{poti}
U(x)=\frac{a|x|^c}{c},
\end{equation}
with $a>0$ and $c\ge 2$, $M$ is the
particle mass, $\gamma$ the friction coefficient, and $Y_{\alpha}(t)$
represents a stationary white L{\'e}vy noise with L{\'e}vy index $\alpha$
($1\le\alpha\le 2$).

We employ the white L{\'e}vy noise $Y_{\alpha}(t)$ such that the process
\begin{equation}
L(\Delta t)=\int_t^{t+\Delta t}Y_{\alpha}(\tau)d\tau,
\end{equation}
which is the time integral over an increment $\Delta t$ is an $\alpha$-stable
process with stationary independent increments. Restricting ourselves to
symmetric L{\'e}vy laws, this implies the characteristic function
\begin{equation}
\hat{p}_L(k,\Delta t)=\exp\left(-D|k|^{\alpha}\Delta t\right),
\end{equation}
The constant $D$ in this description has the meaning of the intensity of
the Langevin source, and $[D]={\rm cm}^{\alpha}/\sec$.

In figure \ref{noise}, we show realisations of white L{\'e}vy noises for
various values of $\alpha$. The sharply pronounced `outliers', due to the
long-tailed nature of the L{\'e}vy stable distribution, are distinct, in
comparison to the Gaussian case $\alpha=2$.

\subsubsection{Fractional Fokker-Planck equation}

The Langevin equation (\ref{langevin}) is still of the Markov-type, and it
is therefore fairly straightforward to show that the corresponding
fluctuation-averaged (deterministic) description is given in terms of the
space-fractional Fokker-Planck equation (\ref{ffpe})
\cite{seshadri,fogedby,chechkin}. In what follows, we derive solution of
equation (\ref{ffpe}) for the $\delta$-initial condition
\begin{equation}
\label{delta}
f(x,0)=\delta(x).
\end{equation}

The space-fractional derivative $\partial^{\alpha}/\partial|x|^{\alpha}$ 
occurring in the fractional Fokker-Planck equation (\ref{ffpe}) is called
the Riesz fractional derivative, defined through
\begin{equation}
\label{riesz1}
\frac{d^{\alpha}f}{d|x|^{\alpha}}=\left\{\begin{array}{ll}
-\frac{D_{+}^{\alpha }f+D_{-}^{\alpha }f}{2\cos(\pi\alpha/2)}, & \alpha\neq 1\\ 
-\frac{d}{dx}Hf, & \alpha=1
\end{array}
\right. ,
\end{equation}
where we use the following abbreviations:
\begin{equation}
\label{riesz2}
({\bf D}_{+}^{\alpha }f)(x)=\frac{1}{{\Gamma (2-\alpha)}}\frac{d^{2}}{d
x^{2}}\int_{-\infty }^{x}\frac{f(\xi ,t)d\xi }{(x-\xi)^{\alpha-1}},
\end{equation}
and
\begin{equation}
\label{riesz2a}
({\bf D}_{-}^{\alpha }f)(x)=\frac{1}{{\Gamma (2-\alpha )}}\frac{d^{2}}{dx^{2}%
}\int_{x}^{\infty }\frac{f(\xi ,t)d\xi }{(\xi-x)^{\alpha -1}} 
\end{equation}
for, respectively, the left and right Riemann-Liouville derivatives
($1\leq \alpha<2$); and \cite{mainardi}
\begin{equation}
({\bf H}f)(x)=\frac{1}{\pi }\int_{-\infty }^{\infty }\frac{f(\xi
)d\xi }{x-\xi }
\end{equation}
is the Gilbert transform operator. The definitions for $\partial^{\alpha}/
\partial|x|^{\alpha}$ demonstrate the strongly non-local property of the
space-fractional Fokker-Planck equation, i.e., strong correlations in $x$.

\subsubsection{Rescaling of the dynamical equations}

Passing to dimensionless variables 
\begin{equation}
x'=x/x_{0}, \quad t'=t/t_0,
\end{equation}
with
\begin{equation}
x_{0}=\left({\frac{{MD\gamma}}{a}}\right)^{1/(c-2+\alpha)},\quad t_0=
\frac{{x_{0}^{\alpha }}}{D},
\end{equation}
the starting equations take the form (we omit primes below) 
\begin{equation}
\label{langevin1}
\frac{dx}{dt}=-\frac{d{U}}{dx}+Y_{\alpha}(t)
\end{equation}
instead of the Langevin equation (\ref{langevin}), and
\begin{equation}
\label{ffpe1}
\frac{\partial f(x,t)}{\partial t}=\frac{\partial}{\partial x}\frac{d{U}}{dx}
f+\frac{\partial^{\alpha}f}{\partial |x|^{\alpha}}
\end{equation}
instead of the fractional Fokker-Planck equation (\ref{ffpe}), and
\begin{equation}
\label{poti1}
U(x)=\frac{|x|^c}{c}.
\end{equation}

\subsection{Starting equations in Fourier space}

If $\hat{f}(k,t)$ denotes the characteristic function (CF), i.e., Fourier
transform of $f(x,t)$, we write
\begin{equation}
f(x,t)\div\hat{f}(k,t),
\end{equation}
where we use the sign $\div$ to denote a Fourier transform pair. Since 
\cite{samko}
\begin{equation}
({\bf D}_{\pm }^{\alpha }f)(x,t)\div (\mp ik)^{\alpha }\hat{f}(k,t),
\end{equation}
and
\begin{equation}
({\bf H}f)(x,t)\div i{\rm sign}(k)\hat{f}(k,t),
\end{equation}
we obtain
\begin{equation}
\frac{\partial^{\alpha}f}{\partial|x|^{\alpha}}\div -|k|^{\alpha}\hat{f}(k,t),
\end{equation}
for all $\alpha $'s. The equivalent of the fractional Fokker-Planck equation
(\ref{ffpe1}) for the CF then follows immediately,
\begin{equation}
\label{ffpef}
\frac{\partial\hat{f}}{\partial t}+|k|^{\alpha }\hat{f}={\sf U}_{k}\hat{f},
\end{equation}
with the initial condition
\begin{equation}
\hat{f}(k,t=0)=1,
\end{equation}
and the normalisation
\begin{equation}
\hat{f}(k=0,t)=1.
\end{equation}
The external potential $U(x)$ turns into the linear differential
operator in $k$,
\begin{eqnarray}
\nonumber
{\sf U}_{k}\hat{f}&=&\int_{-\infty}^{\infty}e^{ikx}\frac{\partial}{\partial
x}\left(\frac{dU}{dx}f\right)dx\\
&=&-ik\int_{-\infty }^\infty e^{ikx}{\rm sign}(x)|x|^{c-1}f(
x,t)dx.
\end{eqnarray}
Next, making use the following `inverse' expressions
\begin{equation}
(\pm ix)^\alpha f(x)\div ({\bf D}_{\pm}^\alpha\hat{f})(k),
\end{equation}
and
\begin{equation}
-i({\rm sign}(x)f(x)\div (H\hat{f})(k). 
\end{equation}
we obtain the explicit expression for the external potential operator,
\begin{equation}
{\sf U}_k\hat{f}=\left\{\begin{array}{ll}
\frac{k}{2\cos(\pi c/2)}\left(D_{+}^{c-1}-D_{-}^{c-1}\right)\hat{f}, &
c\neq 3,5,7,\ldots\\[0.4cm]
(-1)^mk\frac{d^{2m}}{dk^{2m}}H\hat{f}, & c=3,5,7,\ldots
\end{array}\right.
\end{equation}
Note that for the even potential exponents $c=2m+2$ , $m=0,1,2,\ldots$,
we find the simplified expression
\begin{equation}
{\sf U}_k=(-1)^{m+1}k\frac{\partial^{2m+1}}{\partial k^{2m+1}}. 
\end{equation}
We see that the force term can be written in terms of fractional derivatives
in Fourier space, and therefore it is not straightforward to calculate even
the stationary solution of the fractional Fokker-Planck equation (\ref{ffpe1})
in the general case $c\notin\mathbb{N}$. In particular, in this latter case,
the non-local equation (\ref{ffpe1}) in $x$-space translates into a non-local
equation in $k$-space, where the non-locality shifts from the diffusion to
the drift term.

\section{Analytical results}

In the preceding section, we discussed some elementary properties of the
space-fractional Fokker-Planck equation for LFs, in particular, we pointed
out the spatially non-local character of equation (\ref{ffpe1}), and its
Fourier space counterpart (\ref{ffpef}). In this section, we determine
the analytical solution of the fractional Fokker-Planck equation. We
start with the exactly solvable stationary quartic Cauchy oscillator,
to demonstrate
directly the occurring bimodality, and then move on to the general case.
The major results will be the determination of $n$-modality, finite
variance, and the parametric dependence of the associated bifurcations.

\subsection{The stationary quartic Cauchy oscillator}

Let us first regard the case of the stationary quartic potential with $c=4$
for the Cauchy-LF with $\alpha=1$, i.e., the solution of the equation
\begin{equation}
\frac{d}{dx}x^3f_{\rm st}(x)+\frac{d}{d|x|}f_{\rm st}(x)=0,
\end{equation}
or,
\begin{equation}
\frac{d^3\hat{f}_{\rm st}(k)}{dk^3}={\rm sign}(k)|k|\hat{f}_{\rm st}
(k)
\end{equation}
in Fourier space. Its solution is
\begin{equation}
\hat{f}_{\rm st}(k)=\frac{2}{\sqrt{3}}\exp\left(-\frac{|k|}{2}\right)\cos\left(
\frac{\sqrt{3}|k|}{2}-\frac{\pi}{6}\right),
\end{equation}
whose inverse Fourier transform results in the simple analytical form
\begin{equation}
\label{cauchystat}
f_{\rm st}(x)=\frac{1}{\pi(1-x^2+x^4)}.
\end{equation}
As shown in figure \ref{fig0}, this solution has {\em two global maxima\/} at
$x_{\rm max}=\pm 1/\sqrt{2}$ apart from a local minimum at the origin (the
position of the initial condition, that is) and its variance
\begin{equation}
\langle x^2\rangle=1
\end{equation}
is {\em finite\/}, due to the long-tail asymptotics $f_{\rm st}(x)\sim x^{-4}$.
These two distinct properties for LFs will turn out to be a central theme in
the remainder of this work.

\subsection{Formal solution of equation (\protect\ref{ffpe1})}

Returning to the general case, we rewrite equation (\ref{ffpef}) in the
equivalent integral form,
\begin{equation}
\label{conv}
\hat{f}(k,t)=\hat{p}_{\alpha}(k,t)+\int_{0}^{t}d\tau \;\hat{p}_{\alpha}
(k,t-\tau ){\sf U}_{k}\hat{f}(k,\tau)
\end{equation}
where 
\begin{equation}
\hat{p}_{\alpha }(k,t)=\exp(-|k|^{\alpha}t)
\end{equation}
is the CF of a free (homogeneous) LF. This relation follows from equation
(\ref{ffpef}) via formally treating it as a non-homogeneous linear differential
equation of first order, where ${\sf U}_k$ plays the role of the
non-homogeneity.
Then, (\ref{ffpef}) is obtained from variation of constants. (Differentiate
equation (\ref{conv}) to return to (\ref{ffpef}).)

Equation (\ref{conv}) can be solved formally by iterations: Let
\begin{equation}
\hat{f}^{(0)}(k,t)=\hat{p}_{\alpha}(k,t),
\end{equation}
then
\begin{equation}
\hat{f}^{(1)}(k,t)=\hat{p}_{\alpha}(k,t)+\int_0^td\tau \hat{p}_{\alpha}(k,t
-\tau){\sf U}_k\hat{f}^{(0)}(k,\tau),
\end{equation}
\begin{eqnarray}
\nonumber
\hat{f}^{(2)}(k,t)=\hat{p}_{\alpha}(k,t)+\int_0^td\tau \hat{p}_{\alpha}(k,t
-\tau){\sf U}_k\hat{p}_{\alpha}(k,\tau)+\\
\int_0^td\tau\int_0^{\tau}d\tau' \hat{p}_{\alpha}(k,t-\tau){\sf U}_k
\hat{p}_{\alpha}(k,\tau-\tau'){\sf U}_k\hat{p}_{\alpha}(k,\tau'),
\end{eqnarray}
etc. Invoking the definition of the convolution,
\begin{equation}
A*B=\int_{0}^{t}d\tau A(t-\tau)B(\tau)=\int_{0}^{t}d\tau A(\tau)B(t-\tau),
\end{equation}
and using
\begin{equation}
A*B*C=(A*B)*C=A*(B*C),
\end{equation}
we arrive at the formal solution
\begin{equation}
\label{conv1}
\hat{f}(k,t)=\sum_{n=0}^{\infty }\hat{p}{_{\alpha}}(*{\sf U}_{k}\hat{p}
_{\alpha})^n,
\end{equation}
This procedure is analogous to perturbation theory, ${\sf U}_k\hat{f}$
playing the role of the interaction term, see, for instance, reference
\cite{balescu}, chapter 16.

Applying a Laplace transformation,
\begin{equation}
\widetilde{\widehat{f}}(k,s)=\int_{0}^{\infty }dt\,\exp(-st)\widehat{f}(k,t),
\end{equation}
to equation (\ref{conv1}), we obtain
\begin{equation}
\label{conv2}
\widetilde{\widehat{f}}(k,s)=\widetilde{\widehat{p}}_{\alpha }(k,s)+
\widetilde{\widehat{p}}_{\alpha }(k,s){\bf U}_{k}\widetilde{\widehat{f}}
(k,s),
\end{equation}
where 
\begin{equation}
\widetilde{\widehat{p}}_{\alpha}(k,s)=\frac{1}{s+k^{\alpha }}
\end{equation}
is the Fourier--Laplace transform of the homogeneous L{\'e}vy stable PDF.
Thus, we obtain the equivalent of solution (\ref{conv1}) in $(k,s)$-space:
\begin{equation}
\label{conv3}
\widetilde{\widehat{f}}(k,s)=\sum_{n=0}^{\infty }\left[ \widetilde{
\widehat{p}}_{\alpha }(k,s){\sf U}_k\right] ^{n}\widetilde{\widehat{p}}
_{\alpha }(k,s).
\end{equation}
This iterative construction scheme for the solution of the fractional
Fokker-Planck equation will turn out to be useful below.

\subsection{Existence of a bifurcation time.}

For the case of the unimodal initial condition $f(x,0)=\delta(x)$ we now
proof the existence of a finite bifurcation time $t_{12}$ for the turnover
from unimodal to bimodal PDF. At this time, the curvature at the origin will
vanish, i.e., be an inflection point:
\begin{equation}
\label{bif}
\left.\frac{\partial^2f}{\partial x^2}\right|_{x=0,t=t_{12}}=0 
\end{equation}
Introducing
\begin{equation}
\label{bif1}
J(t)=\int_{0}^{\infty }dkk^{2}\hat{f}(k,t),
\end{equation}
equation (\ref{bif}) is equivalent to (note that the CF is an even function) 
\begin{equation}
J(t_{12})=0.
\end{equation}

The bifurcation can now be obtained from the iterative solution (\ref{conv3});
we consider the specific case $c=4$. From the first order approximation
\begin{equation}
\widetilde{\widehat{f}}_1(k,s)=\frac{1}{s+k^{\alpha}}\left(1+{\sf U}_k\frac{1}
{s+k^{\alpha}}\right),
\end{equation}
and we have 
\begin{equation}
{\sf U}_k=k\frac{\partial^3}{\partial k^3}.
\end{equation}
Combining these two expressions, we produce
\begin{eqnarray}
\nonumber
\widetilde{\widehat{f}}_1(k,s)=\frac{1}{s+k^{\alpha}}+\alpha(\alpha
-1)(2-\alpha )\frac{k^{\alpha-2}}{{(s+k^{\alpha})^3}}\\
+6\alpha^2(\alpha-1)\frac{k^{2\alpha-2}}{(s+k^{\alpha })^4}
-6\alpha ^3\frac{{k^{3\alpha -2}}}{{(s+k^\alpha )^5}},
\end{eqnarray}
or, after inverse Laplace transformation, 
\begin{eqnarray}
\nonumber
\hat{f}_1(k,t)= 
e^{-k^{\alpha }t}&&\left\{1-\frac{\alpha^3}{4}t^4k^{3\alpha
-2}+\alpha ^{2}(\alpha -1)t^{3}k^{2\alpha -2}\right.\\
&&\left.+\alpha(\alpha-1)(2-\alpha)\frac{t^2}{2}k^{\alpha-2}\right\} .
\end{eqnarray}
The first approximation to the bifurcation time $t_{12}$ is then determined
via equation (\ref{bif1}), i.e., we calculate
\begin{equation}
\int_0^{\infty}dk\;k^2\hat{f}_1(k,t_{12}^{(1)})=0,
\end{equation}
to obtain
\begin{equation}
t_{12}^{(1)}=\left(\frac{4\Gamma(3/\alpha)}{3(3-\alpha)\Gamma(1/\alpha)}
\right)^{\alpha/(2+\alpha)}.
\end{equation}
In figure \ref{fig2}, we show the dependence of this first approximation
$t_{12}^{(1)}$ as a function of the L{\'e}vy index $\alpha$ (dashed line),
in comparison to the values determined from the numerical solution of the
fractional Fokker-Planck equation (\ref{ffpe1}) shown as the dotted line.
The second order iteration for the PDF, $\hat{f}_{2}(k,t)$, can be obtained
with {\tt maple6}, from which in turn the second approximation for the
bifurcation time is found in analogy to above procedure. The result is
displayed as the full line in figure \ref{fig2}. The two approximative
results show in fact surprisingly good agreement with the numerical
result of the full PDF.

\subsection{Proof of non-uni-modality of stationary solution for $c>2$}

In this subsection we demonstrate that the stationary solution of the
kinetic equation (\ref{ffpe1}) has a non-unimodal shape. For this purpose,
we use an alternative expression for the fractional Riesz derivative
(compare, e.g., reference \cite{samko}), 
\begin{eqnarray}
\nonumber
\frac{d^{\alpha}f(x)}{d|x|^{\alpha}}&\equiv&\Gamma(1+\alpha)\frac{\sin(\alpha
\pi/2)}{\pi}\\
&&\hspace*{-0.6cm}
\times\int_0^\infty d\xi\frac{f(x+\xi)-2f(x)+f(x-\xi )}{\xi^{1+\alpha}}
\label{riesz3}
\end{eqnarray}
valid for $0<\alpha<2$. This definition can be obtained by `regularisation'
of our definitions (\ref{riesz1}) to (\ref{riesz2a}) for $\alpha \neq 1$;
however, expression (\ref{riesz3}) is also valid at $\alpha =1$. In the
stationary state ($\partial f/\partial t=0$), we get from equation
(\ref{ffpe1}):
\begin{equation}
\frac{d}{dx}\left({\rm sgn}(x)|x|^{c-1}f_{\rm st}\right)+
\frac{d^{\alpha}f_{\rm st}}{d|x|^{\alpha}}=0.
\end{equation}
Thus, it follows that at $c>2$ (strict inequality) 
\begin{equation}
\left.\frac{d^{\alpha}f_{\rm st}}{d|x|^{\alpha}}\right|_{x=0}=0,
\end{equation}
or, from definition (\ref{riesz3}) and taking into account that $f_{\rm st}
(x)$ is an even function, 
\begin{equation}
\label{cond}
\int_0^{\infty }d\xi\frac{f_{\rm st}(\xi)-f_{\rm st}(0)}{\xi^{1+\alpha}}=0. 
\end{equation}
From this latter relation, we can immediately obtain proof the non-uni-modality
of $f_{\rm st}$, which we produce in two steps:

(1) If we assume that the stationary PDF $f_{\rm st}(x)$ is unimodal, then due
to the symmetry $x\rightarrow -x$, it necessarily has one global maximum at
$x=0$. In this case the integrand in equation (\ref{cond}) must be negative,
and therefore contradicts equation (\ref{cond}). Therefore, $f_{\rm st}(x)$
is non-unimodal.

(2) We can in addition exclude $f(0)=0$, as in this case the integrand
will be positive, which is again in contradiction with equation (\ref{cond}).

Since $f(x)\rightarrow 0$ at $x\rightarrow\infty$, basing on statements (1) and
(2), one may conclude that the simplest situation is such that $\xi_{0}>0$
exists with the property
\begin{equation}
\int_{\xi_0}^{\infty}d\xi\frac{f(\xi)-f(0)}{\xi^{1+\alpha}}<0,
\end{equation}
and,
\begin{equation}
\int_0^{\xi_0}d\xi\frac{f(\xi)-f(0)}{\xi^{1+\alpha}}>0, 
\end{equation}
i.e., the condition for two-hump stationary PDF for all $c>2$. At intermittent
times, however, we will show that also a tri-modal state may exist.

\subsection{Power-law asymptotics of stationary solutions for $c\geq 2$,
and finite variance for $c>2$}

We now derive the power-law asymptotics of the stationary PDF $f_{\rm st}(x)$
for external potentials of the form (\ref{superharm}) with general $c\ge 2$.
To this end, we note that at $x\rightarrow +\infty$, it is reasonable to
assume
\begin{equation}
D_{-}^{\alpha}f_{\rm st}\ll D_{+}^{\alpha}f_{\rm st},
\end{equation}
since the region of integration for the right-side Riemann-Liouville
derivative $({\bf D}_-^{\alpha}f_{\rm st})(x)$, $(x,\infty)$, is much 
smaller than the region of integration for the left-side derivative
$({\bf D}_+^{\alpha}f_{\rm st})(x)$, $(-\infty ,x)$, in which also the
major portion of $f_{\rm st}(x)$ is located. Thus, at large $x$ we get
for the stationary state, 
\begin{equation}
\frac{d}{d{x}}\left(\frac{dU}{dx}f_{\rm st}\right)-\frac{1}{2\cos(\pi\alpha/2)}
\frac{d^2}{dx^2}\int_{-\infty }^x\frac{f_{\rm st}(\xi)d\xi}{(x-\xi)^{a-1}}\cong 0.
\end{equation}
Equivalently, this corresponds to the approximative equality
\begin{equation}
\label{cond1}
x^{c-1}f_{\rm st}(x)\cong\frac{1}{2\cos(\pi\alpha/2)}\frac{d}{dx}\int_{-
\infty}^x\frac{
f_{\rm st}(\xi)d\xi}{(x-\xi)^{a-1}}.
\end{equation}
We are seeking asymptotic behaviours of $f_{\rm st}(x)$ in the form
$f(x)\approx C_{1}/x^{\mu}$ ($x\to +\infty$, $\mu>0)$. After integration
of relation (\ref{cond1}), we find
\begin{equation}
\frac{2C_1\cos(\pi\alpha/2)\Gamma(2-\alpha)}{-\mu+c}x^{-\mu+c}\cong
\int_{-\infty}^x\frac{f_{\rm st}(\xi)d\xi}{(x-\xi)^{a-1}}.
\end{equation}
The integral on the right hand side can be approximated through
\begin{equation}
\frac{1}{x^{\alpha-1}}\int_{-\infty}^xf_{\rm st}(\xi)d\xi\cong\frac{1}{x^
{\alpha-1}}\int_{-\infty}^{\infty}f_{\rm st}(\xi)d\xi=\frac{1}{x^{\alpha-1}}.
\end{equation}
Thus, we identify the powers of $x$ and the prefactor, with the results
\begin{equation}
\mu =\alpha +c-1
\end{equation}
and 
\begin{equation}
C_1=\frac{\sin(\pi\alpha/2)\Gamma(\alpha)}{\pi}.
\end{equation}
By symmetry of the PDF we therefore recover the general asymptotic form 
\begin{equation}
\label{as}
f(x)\approx\frac{\sin(\pi\alpha/2)\Gamma(\alpha)}{\pi|x|^{\mu}},\,\,x\to
+\infty
\end{equation}
for all $c\geq 2$. This result is remarkable, for various reasons:

(i) despite the approximations involved, the asymptotic form (\ref{as}) for
arbitrary $c\ge 2$ matches exactly previously obtained forms, such as
the exact analytical result for the harmonic LF (linear L{\'e}vy
oscillator), $c=2$ reported in reference \cite{sune}; the result for the
quartic L{\'e}vy oscillator with $c=4$ discussed in references
\cite{chechkin,chechkin2}; and the case of even power-law exponents
$c=2m+2$ ($m\in\mathbb{N}_0$) given in reference \cite{chechkin}.

(ii) The prefactor $C_1$ is independent of the potential exponent
$c$; in this sense, $C_1$ is `universal'.

(iii) For each value $\alpha$ of the L{\'e}vy index the 'critical'
value
\begin{equation}
c_{\rm cr}=4-\alpha
\end{equation}
exists such that at $c<c_{\rm cr}$ the variance $<x^2>$ is infinite,
whereas at $c>c_{\rm cr}$ the variance is {\em finite}.

(iv) We have found a fairly simple trick to construct stationary solutions
at large $x$ in the form of inverse power series.

\section{Numerical results}

In this section, we show results for the PDF of the fractional Fokker-Planck
equation (\ref{ffpe1}) obtained via two different numerical techniques, one
being the Gr{\"u}nwald-Letnikov method, which is based on an iterative
solution of the deterministic dynamical equation (\ref{ffpe1}) by replacing
the Riesz fractional derivative with Gr{\"u}nwald-Letnikov operators; the other
being the Langevin method, in which the Langevin equation with L{\'e}vy
noise, equation (\ref{langevin}) is integrated numerically. Both methods
lead to analogous results, and they also produce results for the PDF which
are in perfect agreement with above analytical results. We present the
numerical results in two subsections, devoted to the two numerical methods.
These methods themselves are discussed in the appendix.

\subsection{Results from the Gr{\"u}nwald-Letnikov method}

\subsubsection{Tri-modal transient state at $c>4$}

Before, we have proved the existence of a bimodal stationary state for the
quartic ($c=4$) L{\'e}vy oscillator. This bimodality emerges as a bifurcation
at a critical time $t_{\rm cr}$, at which the curvature at the origin vanishes.
This scenario is replaced for $c>4$, as displayed in figure \ref{fig3}.
Thus, there obviously exist two timescales, the critical time for the 
emergence of the two off-centre maxima, which are characteristic for
the stationary state; and a second one, which is designated to the
relaxation of the initial central hump, i.e., the decaying initial
distribution $f(x,0)=\delta(x)$. The formation of the two off-centre
humps while the central one is still present, is detailed in figure
\ref{fig4}.
This existence of a transient tri-modal state was found to be typical for
all $c>4$.

In figures \ref{fig5} and \ref{fig6}, we show additional details of the
tri-modal state. Thus, figure \ref{fig5} depicts a bifurcation diagram
for the abovely shown process; the initial mono-modal PDF bifurcates to
a tri-modal one, before it finally turns over to bimodality. In figure
\ref{fig6}, these two turnover times are displayed as function of the
L{\'e}vy index $\alpha$. Clearly, there is always a gap between these
two time scales, leaving the intermittent time for the tri-modal state,
and for $\alpha\to 2$, this region shrinks, both curves converge to
infinity, as in the regular Gaussian case no such bifurcation exists.

\subsubsection{Phase diagrams for $n$-modal states}

Above findings can be put in context with the purely bimodal case
discussed earlier. A convenient way of displaying the $n$-modal
character of the PDF in the presence of a superharmonic external
potential of the type (\ref{poti1}) is the phase diagram
shown in figure \ref{fig7}.
There, we summarise the findings that for $2<c\le 4$ the bifurcation
occurs between the initial mono-modal and the stationary bimodal PDF
at a finite critical time, whereas for $c>4$, a transient tri-modal
state exists. Moreover, we also include the shaded region, in which
$c$ is too small to ensure a finite variance. Complementarily, in 
figure \ref{fig8}, the temporal domains of the $n$-modal states
are graphed, and the solid lines separating these domains correspond
to the critical timescales. Again, the transient nature of the tri-modal
state is distinct.

\subsection{Langevin method}

As explained in the appendix, this method directly integrates the Langevin
equation for white L{\'e}vy noise, the latter being portrayed in figure
\ref{noise}. Typical results for the sample paths under the influence of
an external potential (\ref{poti1}) with increasing superharmonicity are
shown in figure \ref{fig9} in comparison to the Brownian case (i.e., white
Gaussian noise). For increasing the external exponent $c$, the long excursions,
which are typical for homogeneous LFs are increasingly suppressed. In
the harmonic case $c=2$
still present (in this case, the variance is diverging), they are clearly
confined for $c>2$. Note the comparable ordinate windows in comparison to
the significantly different scale in the homogeneous case of figure
\ref{noise}. For all displayed cases, however, the qualitative
behaviour of the noise under the external potential is
different from the Brownian noise even in this case of strong
confinement. In the same figure, we also show the curvature of the
external potential. Additional investigations have shown that the
maximum curvature is always very close to the positions of the
two maxima, leading us to conjecture that they are in fact identical.

The latter observation is further investigated in figure \ref{fig10}. On
a linear scale, the potential well and its curvature are compared to the
stationary PDF, clearly demonstrating the proximity of maximum curvature
and the two maxima.
Figure \ref{fig10} also corroborates on the basis of the Langevin method the 
asymptotic inverse power-law behaviour derived in equation (\protect\ref{as}).

Finally, in figure \ref{fig11}, we display the time evolution of the PDF
in the three different modality-regimes according to figure \ref{fig8}.
The comparatively noisy result is due to a small number of trajectories
used for the statistical average, due to the rather computation intensive
program.

\section{Conclusions}

By combining analytical and numerical results, we discussed LFs in a
superharmonic external potential of power $c$. Depending on the magnitude
of this exponent $c$, different regimes could be demonstrated. Thus, for
$c=2$, the character of the L{\'e}vy noise imprinted on the process,
is not changed by the external potential: the resulting PDF has L{\'e}vy
index $\alpha$, the same as the noise itself, and will thus give rise to
a diverging variance at all times. Conversely, for $c>2$, the variance
becomes finite if only $c>c_{\rm cr}=4-\alpha$. This is due to the fact that
the PDF leaves the class of L{\'e}vy stable PDFs and acquires an inverse
power-law asymptotic behaviour with power $\mu=\alpha+c-1$. Obviously,
moments of higher order still diverge.
Apart from the finite variance, the PDF is distinguished by the
observation that it bifurcates from the initial mono-modal to a stationary
bimodal state. If $c>4$, there exists a transient tri-modal state.
This richness of the PDF both during relaxation and at stationarity,
depending on a competition between L{\'e}vy noise and steepness of the
potential contrasts the universal approach to the Boltzmann equilibrium,
solely defined by the external potential, in classical diffusion.

It may be speculated what the exact kinetic reason for the occurrence of
the multiple humps is. Due to the observation that the non-transient humps
seem to coincide with the positions of maximum curvature of the external
potential, one may conclude that the potential at these points changes
almost abruptly (especially for larger $c$) from a rather flat to a very
steep slope, towards which the random walker is driven by the diffusivity.
I.e., in comparison to the harmonic and subharmonic cases, the restoring
force close to the origin is comparably weak.

The different regimes for $c>2$ can be classified in terms of critical
quantities, in particular, the bifurcation time(s) $t_{\rm cr}$ and the
critical external potential exponent $c_{\rm cr}$. LFs in superharmonic
potentials can then be conveniently represented by phase diagrams on the
$(c,\alpha)$ and $(c,t_{\rm cr})$ plains.

The numerical solution of both the fractional Fokker-Planck equation in
terms of the Gr{\"u}nwald-Letnikov scheme used to find a discretised
approximation of the fractional Riesz operator shows reliable convergence,
as corroborated by direct solution of the corresponding Langevin equation.

Our findings have underlined the statement that the properties of LFs,
in particular under non-trivial boundary conditions or in an external
potential are not fully understood. The general difficulty, which hampers
a similarly straightforward investigation like in the regular Gaussian or
the subdiffusive cases, is connected with the strong spatial correlations
of the problems, manifested in the integrodifferential nature of the
Riesz fractional operator.  For this reason it is already non-trivial to
determine the stationary solution of the process. We expect, by the fact
that diverging fluctuations appear to be relevant in physical systems, a
range of yet unknown properties of LFs remain to be discovered.

\acknowledgments

The authors thank F. Mainardi and R. Gorenflo for stimulating discussions
and G. Voitsenya for help in numerical simulation. This work is supported by
the INTAS Project 00 - 0847.

\begin{appendix}

\section{Numerical solution methods}

In this appendix, we briefly review the numerical techniques, which were
used in this work to determine the PDF from the fractional Fokker-Planck
equation (\ref{ffpe1}) and the Langevin equation (\ref{langevin1}).

\subsection{Numerical solution of the fractional Fokker-Planck equation
(\protect\ref{ffpe1}) via the Gr{\"u}nwald-Letnikov method}

From a mathematical point of view, the fractional Fokker-Planck equation
(\ref{ffpe1}) is an first-order partial differential equation in time, and
of non-local, integro-differential kind in the position co-ordinate $x$.
It can be solved numerically via an efficient discretisation scheme following
Gr{\"u}nwald and Letnikov \cite{podlubny,gorenflo,gorenflo1}.

Let us designate the force component on the right hand side of equation
(\ref{ffpe1}) as 
\begin{equation}
\label{frce}
\bar{F}(x,t)\equiv\frac{\partial}{\partial x}\left(\frac{dU}{dx}f\right);
\end{equation}
and the diffusion part as 
\begin{equation}
\label{diffu}
\bar{D}(x,t)\equiv\frac{\partial^{\alpha}f}{\partial |x|^{\alpha}}.
\end{equation}
With these definitions, we can rewrite equation (\ref{ffpe1}) in terms of
a discretisation scheme as
\begin{equation}
\label{discrete}
\frac{{f_{j,n+1}-f_{j,n}}}{{\Delta t}}=\bar{F}_{j,n}+\bar{D}_{j,n}\,,
\end{equation}
where we encounter the term
\begin{equation}
\bar{F}_{j,n}=x_j^{c-2}\left[(c-1)f_{j,n}+x_j\frac{f_{j+1,n}-f_{j-1,n}}{2
\Delta x}\right],
\end{equation}
which is
the force component of the potential $U(x)=|x|^c/c$. Here, $\Delta t$ and
$\Delta x$ are the finite increments in time and position, such that $t_n
=n\delta t$ and $x_j=j\Delta x$, for $n=0,1,\ldots,N$ and $j=0,1,\ldots,J$,
and $f_{j,n}\equiv f(x_j,t_n)$. Due to the inversion symmetry of the kinetic
equation (\ref{ffpe1}), it is sufficient to solve it on the right semi-axis.
In the evaluation of the numerical scheme, we define $x_J$ such that the PDF
in the stationary state is sufficiently small, say, $10^{-3}$, as determined
from the asymptotic form (\ref{as}).

In order to find a discrete time and position expression for the fractional
Riesz derivative in equation (\ref{diffu}), we employ the Gr{\"u}nwald-Letnikov
scheme \cite{podlubny,gorenflo,gorenflo1}, according to which we obtain
\begin{equation}
\label{diffucoff}
\bar{D}_{j,n}=-\frac{1}{2(\Delta x)^{\alpha}\cos(\pi\alpha/2)}\sum_{q=0}^J
\xi_q\left[f_{j+1-q,n}+f_{j-1+q,n}\right]
\end{equation}
where
\begin{equation}
\xi_q=(-1)^q {\alpha \choose q},
\end{equation}
with
\begin{equation}
{\alpha \choose q}=\left\{\begin{array}{ll}
\alpha (\alpha-1)\ldots(\alpha-q+1)/q!, & q>0\\ 1, & q<0
\end{array}\right.,
\end{equation}
and $1<\alpha\le 2$. Note that in the limiting case $\alpha=2$ only three
coefficients differ from zero, namely, $\xi_0=1$, $\xi_1=-2$, and $\xi_2=1$,
corresponding to the standard three-point difference-scheme for the second
order derivative, $d^2g(x_j)/dx^2\approx (g_{j+1}-2g_j+g_{j-1})/(\Delta x)^2$.
In figure \ref{gl}, we demonstrate that with decreasing $\alpha$, an increasing
number of coefficients contribute significantly to the sum in equation
(\ref{diffucoff}). This becomes particularly clear in the logarithmic
representation in the bottom plot of figure \ref{gl}.
We note that the condition
\begin{equation}
\mu\equiv\Delta t/(\Delta x)^{\alpha}<0.5
\end{equation}
is needed to ensure the numerical stability of the discretisation scheme.
In our numerical evaluation, we use $\Delta x=10^{-3}$, and therefore the
associated time increment $\Delta t\sim 10^{-5}\ldots 10^{-6}$, depending
on the actual value of $\alpha$. The initial condition for equation
(\ref{discrete}) is $f_{0,0}=1/\Delta x$.

In figure \ref{gl1}, the time evolution of the PDF is shown together with
the evolution of the force and diffusion components defined by equations
(\ref{frce}) and (\ref{diffu}), respectively. Accordingly, at the initial
stage of the relaxation process, the diffusion component prevails. The 
force term grows in the course of time, until at the stationary state
$\bar{F}\to-\bar{D}$. This is particularly visible at the bottom right
part of figure \ref{gl1}, which corresponds to the stationary bimodal
state shown to the left.

\subsection{Numerical solution of the Langevin equation
(\protect\ref{langevin})}

An alternative way to obtain the PDF is to sample the trajectories determined
by the Langevin equation (\ref{langevin}). To this end, equation
(\ref{langevin1}) is discretised in time according to
\begin{equation}
x_{n+1}=x_n+F(x_n)\Delta t+(\Delta t)^{1/\alpha}Y_{\alpha}(n\Delta t),
\end{equation}
with $t_n=n\Delta t$ for $n=0,1,2,\ldots$, and where the force $F(x_n)$
is the dimensionless force field at position $x_n$. The sequence $\{Y_{\alpha}
(n\Delta t)\}$ is a discrete-time approximation of a white L{\'e}vy noise
of index $\alpha$ with a unit scale parameter. That is, the sequence of
independent random variables possessing the characteristic function
$\hat{p}=\exp\left(-|k|^{\alpha}\right)$. To generate this sequence $\{Y_{
\alpha}(n\Delta t)\}$, we used the method outlined in reference \cite{chego}.

\end{appendix}

\clearpage

\begin{figure}
\caption{Time evolution of the LF-PDF in the presence of the superharmonic
external potential (\protect\ref{superharm}) with $m=1$ (quartic L{\'e}vy
oscillator) and L{\'e}vy index $\alpha=1.2$, obtained from the numerical
solution of the fractional Fokker-Planck equation, using the
Gr{\"u}nwald-Letnikov representation of the fractional Riesz derivative
(full line). The initial condition is a $\delta$-function at the origin.
The dashed lines indicate the corresponding Boltzmann distribution. The
transition from one to two maxima is clearly seen. This picture of the
time evolution is typical for $2<c\leq 4$ where $c$ is defined in equation
(\ref{poti}) below. The corresponding
location of the maximum/maxima as a function of time is shown in figure
\protect\ref{figInt2}.
\label{figInt1}}
\end{figure}

\begin{figure}
\caption{Bifurcation diagrams for the case $m=1.0$, $\alpha=1.2$, corresponding
to the PDF shown in figure \protect\ref{figInt1}. Left: the thick lines show
the location of the maximum, which at the bifurcation time $t_{12}=0.83\pm
0.01$ turns into two maxima. Right: the value of the PDF in the maxima
location (thick line) and the value in the minimum at $x=0$ (thin line).
\label{figInt2}}
\end{figure}

\begin{figure}
\caption{\label{noise}
White L\'{e}vy noises with the L\'{e}vy indexes $\alpha=2,1.7,1.3,1.0$. The
`outliers' are increasingly more pronounced the smaller the L{\'e}vy
index $\alpha$ becomes. Note the different scales on the ordinates.
}
\end{figure}

\begin{figure}
\caption{Stationary PDF (\protect\ref{cauchystat}) of the Cauchy-LF in
a quartic ($c=4$) potential. Two global maxima exist at $x_{\rm max}=\pm
\sqrt{1/2}$, and there is a local minimum at the origin.
\label{fig0}}
\end{figure}

\begin{figure}
\caption{Bifurcation time $t_{12}$ versus L{\'e}vy exponent $\alpha$ at
external potential exponent $c=4.0$. Black dots: bifurcation time deduced
from the numerical solution of the fractional Fokker-Planck equation
(\protect\ref{ffpe1})  using the Gr{\"u}nwald-Letnikov representation of the
fractional Riesz derivative (see appendix). Dashed line: first approximation
$t_{12}^{(1)}$; solid line: second approximation $t_{12}^{(2)}$.
\label{fig2}}
\end{figure}

\begin{figure}
\caption{Time evolution of the PDF governed by the fractional Fokker-Planck
equation (\protect\ref{ffpe1}) in a superharmonic potential
(\protect\ref{superharm}) with exponent $c=5.5$, and for L{\'e}vy index
$\alpha=1.2$; obtained from numerical solution using the  Gr{\"u}nwald-Letnikov
method explained in the appendix. Initial condition is $f(x,0)=\delta(x)$.
The dashed lines indicate the corresponding Boltzmann distribution. The
transitions between $1\rightarrow 3\rightarrow 2$ humps are clearly seen.
This picture of time evolution is typical for $c>4$. On a finer scale, we
depict the transient tri-modal state in figure \protect\ref{fig4}.
\label{fig3}}
\end{figure}

\begin{figure}
\caption{The transition $1\rightarrow 3\rightarrow 2$ from figure
\protect\ref{fig3} on a finer scale ($c=5.5$, $\alpha=1.2$).
\label{fig4}}
\end{figure}

\begin{figure}
\caption{Bifurcation diagrams for the case $c=5.5$ and $\alpha=1.2$
corresponding to figures \protect\ref{fig3} and \ref{fig4}. Left:
positions $x_{\rm max}$ of the maxima (global and local, thick lines);
the thin lines indicate the positions of the minima (at the first
bifurcation time, there is a horizontal tangent at the site of the
two emerging off-centre maxima). The bifurcation times are $t_{13}=0.75\pm
0.01$ and $t_{32}=0.92\pm 0.01$. Right: values of the PDF at the maxima
(thick lines); the thin line indicates the value of the PDF in the minima.
\label{fig5}}
\end{figure}

\begin{figure}
\caption{Bifurcation times $t_{13}$ versus $\alpha $ (lower curve) and 
$t_{32}$ (upper curve) for the external potential exponent $c=5.5$.
\label{fig6}}
\end{figure}

\begin{figure}
\caption{$(c,\alpha)$-map showing different regimes of evolution of the
PDF, and the stationary states. The region with infinite variance is
shaded. The region $c<4$ covers the transition from 1 to 2 humps during
time evolution. For $c>4$, a transition from 1 to 3, and then from 3 to 2
humps occur.  For all $c$'s there are two maxima in the stationary state.
Compare figure \protect\ref{fig8}.
\label{fig7}}
\end{figure}

\begin{figure}
\caption{$(c,t)$-map showing states with different number of maxima, and
the transitions between them during time evolution. Region 1: PDF has 1 hump;
region 2: PDF has 2 humps; region 3: PDF has 3 humps. At $c<4$ there is only
a transition $1\rightarrow 2$, whereas at $c>4$ there are two transitions:
$1\rightarrow 3$ and after the transient tri-modal regime, $3\rightarrow 2$.
\label{fig8}}
\end{figure}

\begin{figure}
\caption{
Left column: the potential energy functions $U=x^{c}/c$, (solid
lines) and their curvatures (dotted lines) for different values of $c$: $c=2$
(linear oscillator), and $c=4,6,8$ (strongly non-linear oscillators). Middle
column: typical sample paths of the Brownian oscillators, $\alpha =2$, with
the potential energy functions shown on the left. Right column: typical
sample paths of the L{\'e}vy oscillators, $\alpha =1$. It is seen that with $c$
increasing the potential walls become steeper, and the L\'{e}vy flights
become shorter; in this sense, they are ''confined''.
\label{fig9}}
\end{figure}

\begin{figure}
\caption{Stationary PDF $f_{\rm st}(x)$ on linear (left) and
double-logarithmic (right) scale, obtained from the Langevin equation for
a) $c=4.0$, b) $c=5.7$, and c) $c=6.5$. The thin lines on the left show the
potential wells and their curvatures. The solid lines on the right show
the asymptotics as given by equation (\protect\ref{as}).
\label{fig10}}
\end{figure}

\begin{figure}
\caption{Time evolution of the PDF $f(x)$ obtained from the Langevin
equation for $c=5.5$, $\alpha =1.2;$ besides the statistical averaging
over different trajectories, a time averaging over a small time interval
has been used to create these figures: a) inside the region of a single
peak, $t<t_{13}$; b) inside the region of three peaks, $t_{13}<t<t_{32}$;
c) inside the region of two peaks,  $t>t_{32}$. \label{fig11}}
\end{figure}

\begin{figure}
\caption{Coefficients
$\xi _{q}$ in Gr\"{u}nwald-Letnikov approximation for different values of the
L\'{e}vy index $\alpha=1.9$, 1.5, and 1.1.
\label{gl}}
\end{figure}

\begin{figure}
\caption{Further details of the Gr\"{u}nwald-Letnikov scheme. Left: Time
evolution of the PDF as obtained by numerical solution of equation
(\protect\ref{discrete}) at $c=4$ and $\alpha=1.2$. Right: Time evolution
of the diffusion component (\protect\ref{diffu}) (thick lines), and the
force term (\protect\ref{frce}) (thin lines).
\label{gl1}}
\end{figure}


\begin{thebibliography}{99}
\bibitem{rem} A more formal form way of writing this Langevin equation is
\[
x(t+dt)-x(t)
=-\frac{1}{m\gamma}\frac{dU(x)}{dx}dt+D^{1/\alpha}Y_{\alpha}(dt).
\]
\end{thebibliography}
\end{document}